\documentclass [12pt]{article}
\def\maj#1{\ifmmode\mbox{\usefont{U}{msb}{m}{n}#1}\else{\usefont{U}{msb}{m}{n}#1}\fi}
\def\v#1{\mathbf{#1}}

\usepackage{graphics,epsfig,graphicx}
\usepackage{color}

\begin{document}

\title{\textbf{Shiva and Kali diagrams for composite quantum particle many-body effects}}
\author{M. Combescot and O. Betbeder-Matibet
\\ \small{\textit{Institut des NanoSciences de Paris,}}\\
\small{\textit{Universit\'e Pierre et Marie Curie, 
CNRS,}}\\
\small{\textit{Campus Boucicaut, 140 rue de Lourmel, 75015 Paris}}}
\date{}
\maketitle

\begin{abstract}
For half a century, Feynman diagrams have provided an enlightening way of representing many-body effects between elementary fermions and bosons. They however are quite inappropriate to visualize fermion exchanges taking place between a large number of composite quantum particles. We propose to replace them by ``Shiva diagrams'' for cobosons made of two fermions and by Shiva-like and ``Kali diagrams'' for cofermions made of three fermions. We also show how these fermion exchanges formally appear in a many-body theory appropriate to composite quantum particles. This theory relies on an operator algebra based on commutators and anticommutators, the usual scalar algebra based on Green functions being valid for elementary bosons or fermions having strict commutation relations, only.
\end{abstract}


\newpage

Although a large amount of physical problems deals with composite quantum particles, all textbooks on many-body physics [1,2] are restricted to elementary fermions or bosons, by lack of appropriate procedures to properly handle fermion exchanges resulting from the Pauli exclusion principle between the fermionic components of these particles. This is why various sophisticated procedures [3-5] were proposed to map the original fermion subspace into a subspace made of effective particles, which are then taken as elementary, these particles interacting through scatterings constructed on the elementary scatterings between fermions, dressed by a certain amount of fermion exchanges [6,7].

Finding these mapping procedures somewhat unsatisfactory, we decided to face the compositeness of the particles in its full complexity. The difficulty is actually double: to properly generate fermion exchanges taking place between a large number of composite particles, but also to cleanly define what must be called interaction \emph{between} such badly defined objects which continuously exchange their undistinguishable fermions. A few years ago, we tackled the simplest problem on composite quantum particles, namely, particles made of just two elementary fermions [8-10]. We called them ``cobosons'', as a contraction of composite bosons. We showed how, through a set of two commutators, we can reach the $2\times 2$ scatterings describing interactions between the fermions of two cobosons, in the absence of fermion exchanges. Through two other commutators, we can reach the $2\times 2$ scatterings for fermion exchanges between two cobosons, in the absence of fermion interaction. By combining these 2$\times 2$ exchanges, it is possible to generate all possible exchanges taking place between an arbitrary number of cobosons. These exchanges are nicely visualized through a set of new diagrams [9,10], that we called ``Shiva'' from the hindu God, due to their multiarm structure. Like Feynman diagrams [1,2], these Shiva diagrams allow us to readily calculate the physical effects they represent, through rather simple and intuitive rules.

The extension of this many-body theory, originally constructed for composite excitons made of one electron and one hole, to composite fermions made of three fermions --- that we are going to call ``cofermions'' --- is not so straightforward. The fact that, instead of ``two-arm" particles, we now have ``three-arm'' particles, leads to a far more complex diagram topology for fermion exchanges. The operator algebra from which these exchanges follow, is also more elaborate. This pushes us to carefully reconsider the reasons for using four commutators in the case of composite particles made of two fermions, in order to possibly extend the procedure to more complicated quantum objects.

The purpose of the present letter is to work out from the very first line, the structure of the many-body theory necessary to describe three-fermion particles --- and more generally $n$-fermion particles. We are going to show that other diagrams, called ``Kali'', are needed, in addition to Shiva-like diagrams, to possibly describe fermion exchanges taking place between cofermions. Also, instead of just two commutators, we now need three commutation relations, namely, two anticommutators plus one commutator, to fully control all fermion exchanges taking place between three-fermion particles. By contrast, the $2\times 2$ scatterings for fermion interactions in the absence of fermion exchanges, are still generated by  two commutation relations only. These are two commutators in the case of cobosons, and one commutator plus one anticommutator in the case of cofermions.

\section{Shiva and Kali diagrams}

\subsection{Fermion exchange between two-fermion particles}

Let us first consider coboson made of  two different fermions $\alpha$ and $\beta$ [9,10]. Two such cobosons can go from states $i$ and $j$ to states $m$ and $n$, under a simple exchange of their fermions $\alpha$, as shown in Fig.1(a). If a third coboson in state $k$ is involved, we get the diagram of Fig.1(b); and so on, as shown by the Shiva diagram for $N$-body exchange of Fig.1(c). For an easy extension to more complicated composite particles, it is of interest to note that the ``star-topology'' (1b) can be replaced by the ``line-topology'' shown in Figs.1(d) or its symmetrical form shown in Fig.1(e). These three diagrams represent exactly the same exchange: Coboson $m$ has the same fermion $\alpha$ as $i$ and the same fermion $\beta$ as $j$. In the same way, the ``star-topology'' (1c) can be replaced by the ``line-topology'' (1f) or its symmetrical form (not represented).

\subsection{Fermion exchange between three-fermion particles}

We now turn to cofermion made of three different fermions $(\alpha,\beta,\gamma)$. In a fermion exchange, such a cofermion can ``explode'' into either (2+1) or (1+1+1) fermions.

If just two cofermions are involved, we can only have the process shown in Fig.2(a), which is identical to the one of Fig.2(b). Indeed, both cofermions explode into (2+1) fermions, the cofermions $m$ and $j$ having only one fermion in common.

If three cofermions are involved, we can have the three cofermions exploding in (2+1) fermions, as in Fig.2(c). We can also have two cofermions exploding into (2+1) fermions and one cofermion exploding into (1+1+1), as in Fig.2(d). Finally, we can have the three cofermions exploding into (1+1+1) fermions: This last case can be represented either in a ``star-topology'' by the diagram of Fig.3(a), or in a`` line-topology'' by the diagram of Fig.3(b).

We see that the diagrams of Fig.2 have a Shiva-like topology, while the diagrams of Fig.3 have a more complex multiarm topology. We are going to call them ``Kali'', from the hindu Goddess, not as kind as Shiva. 

If we now consider exchanges between four cofermions, we can have processes in which two cofermions at least explode into (2+1) fermions. These are shown by the Shiva-like diagrams of Fig.4(a,b,c). We can also have process in which only one cofermion explodes into (2+1) fermions as in the mixed Shiva-Kali diagram of Fig.4(d). Finally, all cofermions can explode into (1+1+1) fermions as in the two different Kali diagrams of Fig.5.

Although somewhat complicated, these diagrams are still quite nice in the sense that the quantities they
represent are readily obtained, as usual, by writing the product of the wave functions of the ``in'' composite particles on the right side and the complex conjugate of those on the left side, with the fermion variables read from the diagram, and by integrating over all dumb fermion variables [9,10].

These Shiva and Kali diagrams represent fermion exchanges taking place between composite quantum particles. Those are  the tricky part of their many-body physics. In addition to these exchanges, composite particles also interact, in a more conventional way, through interactions which exist between their fermionic components. These appear as additional interaction lines between any two composite particles, as shown in Fig.6.

In the next section, we outline how this nicely intuitive diagrammatic representation for composite-particle many-body effects, can be generated from hard algebra.

\section{Many-body formalism for composite quantum particles}

We again concentrate on cobosons made of fermions $(\alpha,\beta)$ and cofermions made of fermions $(\alpha,\beta,\gamma)$. These fermions, which can be electrons with up or down spin, proton, neutron, valence hole, and so on, are assumed to be different; identical fermions, $\alpha\equiv\beta$, like in the case of the semiconductor triplet trion, will be considered elsewhere. The commutation relations between different elementary fermions read as $[a_{\v k_\alpha}^\dag,b_{\v k_\beta}^\dag]_{\eta_{ab}}=0=[a_{\v k_\alpha},b_{\v k_\beta}^\dag]_{\eta_{ab}}$, where $[A,B]_\eta=AB+\eta BA$, while $\eta_{ab}=\pm 1$. For electrons with up and down spins or semiconductor electrons and holes, $\eta_{ab}$ is equal to 1, while for electrons and protons, $\eta_{ab}=-1$. Fortunately, the key equations which control the  many-body physics of composite quantum particles do not depend on these $\eta_{ab}$ since they only appear through $\eta_{ab}^2$, as possible to check. This is why we can, for simplicity, consider that all elementary fermion operators anticommute, even if they correspond to different quantum particles.

We also assume for simplicity, that these composite quantum particles are Hamiltonian eigenstates, in order to form a complete normalized basis for 2-fermion and 3-fermion states. The creation operators for free and correlated fermions are then simply linked by
\begin{equation}
B_i^\dag=\sum_{\v k_\alpha,\v k_\beta}a_{\v k_\alpha}^\dag b_{\v k_\beta}^\dag \,\langle\v k_\beta,\v k_\alpha|i\rangle\ ,
\end{equation}
\begin{equation}
a_{\v k_\alpha}^\dag b_{\v k_\beta}^\dag=\sum_iB_i^\dag\langle i|\v k_\alpha,\v k_\beta\rangle\ ,
\end{equation}
for cobosons, while for cofermions, this link reads
\begin{equation}
F_i^\dag=\sum_{\v k_\alpha,\v k_\beta,\v k_\gamma}a_{\v k_\alpha}^\dag b_{\v k_\beta}^\dag
c_{\v k_\gamma}^\dag \,\langle\v k_\gamma,\v k_\beta,\v k_\alpha|i\rangle\ ,
\end{equation}
\begin{equation}
a_{\v k_\alpha}^\dag b_{\v k_\beta}^\dag c_{\v k_\gamma}^\dag=\sum_iF_i^\dag\langle i|\v k_\alpha,\v k_\beta,\v k_\gamma\rangle\ .
\end{equation}

\subsection{Fermion exchanges between composite particles}

(i) Composite quantum particles made of an even number of fermions are known to behave as bosons, while those which  are made of an odd number, are fermion-like. This shows up through the commutation relation of their creation operators $C_i^\dag$.  It reads
\begin{equation}
[C_m^\dag,C_i^\dag]_{\eta_1}=0\ ,
\end{equation}
with $\eta_1=-1$ for cobosons like $B_i^\dag$ and $\eta_1=+1$ for cofermions like $F_i^\dag$. If we now turn to the commutation relation between destruction and creation operators, we note that  
$[C_m,C_i^\dag]_{\eta_1}$ acting on vacuum gives $\delta_{m,i}\,|v\rangle$ whatever $\eta_1$ is. It however appears as natural to take the same commutation relation for $(C_m^\dag,C_i^\dag)$ and for $(C_m,C_i^\dag)$. This leads us to write [8,10]
\begin{equation}
[C_m,C_i^\dag]_{\eta_1}=\delta_{m,i}-D_{mi}\ ,
\end{equation}
where the operator $D_{mi}$, which differs from zero for composite particles, is such that $D_{mi}|v\rangle=0$. A precise calculation shows that the operator $D_{mi}$ is a sum of products like $a^\dag a$ in the case of two-fermion particles, while it also contains products like $a^\dag b^\dag ba$ for three-fermion particles; and so on for $n$-fermion particles.

(ii) To go further, we note that Eq.(5) leads to
\begin{equation}
\left[[C_m,C_i^\dag]_{\eta_1},C_j^\dag\right]_{\eta_2}=-\eta_1\left[[C_m,C_j^\dag]_{-\eta_1\eta_2},C_i^\dag\right]_{-1}\ ,
\end{equation}
whatever $(\eta_1,\eta_2)$ are.
Consequently, in order to have a $(i,j)$ symmetry, requirement which can seem as physically relevant, we are led to take $\eta_2=-1$ for \emph{both}, cobosons and cofermions. Homogeneity then leads to consider a second commutation relation, which reads as
\begin{equation}
[D_{mi},C_j^\dag]_{-1}=\sum_nC_n^\dag D_{nmij}\ .
\end{equation}
$D_{nmij}$ reduces to a scalar $D_{nmij}^{(0)}$ in the case of two-fermion particles, since $D_{mi}$ is in $(a^\dag a,\,b^\dag b)$ only, while it also contains an operator $D_{nmij}^{(1)}$ in $(a^\dag a$, $b^\dag b$, $c^\dag c)$ in the case of three-fermion particles, due to the presence of operators like $a^\dag b^\dag ba$ in their $D_{mi}$. 

From Eq.(8) acting on $|v\rangle$, it is then possible to show that, for cobosons and cofermions, the scalar part of $D_{nmij}$ is given by
\begin{equation}
D_{nmij}^{(0)}=\left(\delta_{m,i}\,\delta_{n,j}-\eta_1\,\delta_{m,j}\,\delta_{n,i}\right)-
\langle v|C_nC_mC_i^\dag C_j^\dag|v\rangle\ .
\end{equation}
The first term in the RHS of the above equation, just corresponds to the scalar product appearing in the second term, for particles $(m,n)$ and $(i,j)$ taken as elementary. Equation (9) thus shows that $D_{nmij}^{(0)}$ just corresponds to all possible fermion exchanges taking place between \emph{two} composite particles starting in $(i,j)$ states and ending in $(m,n)$ states, i.e., diagram like the one of Fig.1(a) in the case of two-fermion particles and Fig.2(a) in the case of three-fermion particles.

In the case of two-fermion particles [8,10], commutator (8) then reduces to two terms only,
\begin{equation}
[D_{mi},B_j^\dag]_{-1}=\sum_nB_n^\dag\sum_\rho\lambda_\rho\left(^{n\ \,j}_{m\ i}\right)\ ,
\end{equation}
since coboson $i$ can exchange one of its two fermions, $\rho=\alpha$ or $\beta$, with coboson $j$, to give cobosons $m$ and $n$.

For three-fermion particles, cofermion $i$ can exchange one of its three fermions, $\rho=\alpha$, $\beta$ or $\gamma$, with cofermion $j$; but it can also exchange two of its three fermions. Since $\lambda_{\alpha\beta}\left(^{n\ \,j}_{m\ i}\right)$ is nothing but $\lambda_\gamma\left(^{n\ \,i}_{m\ j}\right)$, as seen from Figs.2(a,b), the scalar $D_{nmij}^{(0)}$ contains two sets of three terms, with opposite signs, as the second set corresponds to a double exchange. By analogy with Eq.(10), this leads us to write the commutator (8) as
\begin{equation}
[D_{mi},F_j^\dag]_{-1}=\sum_nF_n^\dag\sum_\rho\left\{\lambda_\rho\left(^{n\ \,j}_{m\ i}\right)-(i\leftrightarrow j)\right\}+D_{mij}^\dag\ ,
\end{equation}
where the operator $D_{mij}^\dag=\sum_nF_n^\dag D_{nmij}^{(1)}$ gives zero when acting on vacuum.

(iii) To get rid of this operator $D_{mij}^\dag$, we need a third commutation relation. The choice between commutator and anticommutator is again made by enforcing a $(j,k)$ symmetry in
\begin{equation}
\left[\left[[F_m,F_i^\dag]_{+1},F_j^\dag\right]_{-1},F_k^\dag\right]_{\eta_3}=-
\left[\left[[F_m,F_i^\dag]_{+1},F_k^\dag\right]_{-\eta_3},F_j^\dag\right]_{+1}\ .
\end{equation}
This requires $\eta_3=+1$. After some algebra, we end with 
\begin{equation}
[D_{mij}^\dag,F_k^\dag]_{+1}=\sum_{p,n}F_p^\dag F_n^\dag\left\{\chi
\left(\begin{array}{ll}p&k\\n&j\\m&i\end{array}\right)\ +\ \mathrm{perm.}\right\}
\end{equation}
where the scalar $\chi\left(\begin{array}{ll}p&k\\n&j\\m&i\end{array}\right)$ corresponds to the Kali diagram shown in Fig.3, with all possible permutations of the indices on the right, namely, the circular permutations which transform $(i,j,k)$ into $(j,k,i)$ and $(k,i,j)$, and the three non-circular permutations which transform $(i,j,k)$ into $(i,k,j)$, $(j,i,k)$ and $(k,j,i)$, the last three terms appearing with an opposite sign.

\subsection{Fermion interactions between composite particles}

(i) We now turn to scatterings induced by fermion interactions. In order to choose between commutator and anticommutator, we can note that $H$ acting on a state made of one of these composite particles plus an arbitrary state $|\psi\rangle$ must give
\begin{equation}
HC_i^\dag|\psi\rangle=E_iC_i^\dag |\psi\rangle+C_i^\dag H|\psi\rangle+\ldots\ ,
\end{equation}
for $(H-E_i)C_i^\dag|v\rangle=0$.  This leads us to consider the commutator of $H$ and $C_i^\dag$ for both, cobosons and cofermions. This commutator, written as [8,10]
\begin{equation}
[H,C_i^\dag]_{-1}=E_iC_i^\dag+V_i^\dag\ ,
\end{equation}
allows to define the ``creation potential'' $V_i^\dag$ of particle $i$. It describes the interactions of this particle with the rest of the system, due to the elementary interactions of its fermionic components. From homogeneity, this operator must read as
$V_i^\dag=\sum_mC_m^\dag V_{mi}^{(1)}$, with $V_{mi}^{(1)}|v\rangle=0$, as obtained from $V_i^\dag
|v\rangle=0$, which readily follows from Eq.(15) acting on vacuum. A precise calculation of this operator shows that it reads as a sum of products like $a^\dag a$.

(ii) To get rid of the operator $V_i^\dag$, we need a last commutation relation. The choice between commutator and anticommutator again follows from $(i,j)$ symmetry in
\begin{equation}
\left[[H,C_i^\dag]_{-1},C_j^\dag\right]_{\eta'_2}=-\eta_1\left[[H,C_j^\dag]_{-\eta_1\eta'_2},C_i^\dag\right]
_{\eta_1}\ .
\end{equation}
This requires $\eta'_2=\eta_1$, i.e., a commutator for cobosons and an anticommutator for cofermions. Homogeneity then leads to write [8,10]
\begin{equation}
[V_i^\dag,C_j^\dag]_{\eta_1}=\sum_{m,n}C_m^\dag C_n^\dag\,\xi\left(^{n\ \,j}_{m\ i}\right)\ ,
\end{equation}
where $\xi\left(^{n\ \,j}_{m\ i}\right)$ is a scalar. This scattering corresponds to interactions between the elementary fermions of the composite particles $(i,j)$, in the absence of fermion exchange, i.e., with $m$ and $i$ made with the same fermions. It is represented by the diagrams of Fig.6.

\subsection{Structure of the key equations}

We see that the scattering between two composite quantum particles, which comes from elementary-fermion interactions, follows from a first commutator between the system H amiltonian and the creation operator of the particle at hand, whatever the particle is, boson-like or fermion-like. Such a commutator generates a ``creation potential'' which describes the interaction of this composite particle with the rest of the system. By taking the commutator or the anticommutator of this creation potential with a second particle creation operator --- the choice depending whether the particles are boson-like or fermion-like --- we can reach the direct scattering of the two composite particles coming from interactions between their elementary fermions, in the absence of fermion exchange.

The scatterings coming from fermion exchanges in the absence of fermion interaction are more subtle to generate. They come from two commutators in the case of cobosons made of 2 fermions. For cofermions made of 3 fermions, we need two anticommutators plus one commutator, while for 4-fermion particles, we need four commutators, and so on\ldots  This set of commutation relations allows us to generate all fermion exchanges taking place in the scalar products of two-particle states, three-particle states, etc\ldots
These exchanges are visualized by multiarm diagrams that we have called Shiva and Kali, the corresponding exchange scatterings being readily calculated from these diagrammatic representations, through fully intuitive rules.

\section{A few simple applications}

Since exchanges between two-fermion particles were extensively studied in our previous works on composite excitons [10], let us end this letter by a few problems involving exchange processes between cofermions made of three fermions.

(i) According to Eqs.(6,11), the scalar product of two-cofermion states is given by
\begin{equation}
\langle v|F_nF_mF_i^\dag F_j^\dag|v\rangle=\left\{\delta_{m,i}\delta_{n,j}-\sum_\rho
\lambda_\rho\left(^{n\ \,j}_{m\ i}\right)\right\}-\{i\leftrightarrow j\}\ ,
\end{equation}
where $\lambda_\rho\left(^{n\ \,j}_{m\ i}\right)$ corresponds to the Shiva-like diagram of Fig.2(a) in which the two cofermions exchange a fermion $\rho$. We readily recover that this scalar product reduces to zero for $(i=j)$ or $(m=n)$, as necessary since $F_i^{\dag 2}|v\rangle=0$, due to Eq.(5). The above equation also shows that the normalization factor of a two-cofermion state $F_i^\dag F_j^\dag|v\rangle$, with $\langle v|F_nF_n^\dag|v\rangle=1$ for $n=(i,j)$, is given by
\begin{equation}
\langle F_jF_iF_i^\dag F_j^\dag|v\rangle=(1-\delta_{i,j})\left[1-\sum_\rho\left\{\lambda_\rho\left(^{j\ j}_{i\ i}\right)-\lambda
_\rho\left(^{j\ i}_{i\ j}\right)\right\}\right]=(1-\delta_{i,j})[1-X_{ij}]\ .
\end{equation}
As for elementary fermions, this normalization factor reduces to zero for $i=j$. However, unlike them, it is not exactly equal to 1 for $i\neq j$, due to the Pauli exclusion principle which generates the exchange term of Eq.(19).
Such a ``moth-eaten effect'', resulting from this Pauli exclusion which enforces the second cofermion to be ``incomplete'', is expected to decrease the elementary fermion value of the scalar product, in the same way as for coboson states. Indeed, as previously shown [11,12], the normalization factor for $N$ identical cobosons 0, namely,
$\langle v|B_0^NB_0^{\dag N}|v\rangle$, reads as $N!\,F_N$, where $F_N$, always smaller than 1, turns exponentially small in the large $N$ limit.

(ii) A similar ``moth-eaten effect'' also exists for the normalization factor of three-cofermion states. By using Eqs.(6,11,13), we find that, for different $(i, j, k)$, this normalization factor reads
\begin{equation}
\langle v|F_kF_jF_iF_i^\dag F_j^\dag F_k^\dag|v\rangle=1-(X_{ij}+X_{jk}+X_{ki})+S_2+S_3+K\ .
\end{equation}

The first term, 1, comes from process in which the three cofermions keep their three fermions, as if these were elementary particles.
In the second term, two cofermions among three are involved in exchanges similar to the ones appearing in Eq.(19), the third cofermion staying unchanged.
The term $S_2$ corresponds to the Shiva-like diagram of Fig.2(d), with two cofermions among three exploding into (2+1) fermions.
The term $S_3$ corresponds to the Shiva-like diagram of Fig.2(c) in which all three cofermions explode into (2+1) fermions. 
Finally, the last term $K$ corresponds to the Kali diagram of Fig.3(a) or Fig.3(b), in which all three cofermions explode into (1+1+1) fermions. 

This scalar product is definitely rather awful. We must however note that  we actually are handling $3\times 3$ fermions on each side, i.e., 18 quantum particles. Thanks to the Shiva and Kali diagrams introduced to visualize composite-particle many-body effects, we can not only understand these exchanges but also classify them in a systematic way, a blind brute force calculation, always possible when the number of cofermions is small, being hardly extendable to larger number of cofermions. 

(iii) We can also study the energy of cofermion states through the Hamiltonian mean value, as we did for cobosons [10,13]. In the two-cofermion subspace, this Hamiltonian mean value reads as
\begin{equation}
\frac{\langle v|F_jF_iHF_i^\dag F_j^\dag|v\rangle}{\langle v|F_jF_iF_i^\dag F_j^\dag|v\rangle}=
[E_i+E_j]+C_{ij}\ .
\end{equation}
The first term corresponds to the energies of the free cofermions $i$ and $j$, while the second term comes from interactions, its precise value reading
\begin{equation}
C_{ij}=\frac{\left[\xi\left(^{j\ j}_{i\ i}\right)-\xi^\mathrm{in}\left(^{j\ j}_{i\ i}\right)\right]-\left[\xi\left(^{j\ i}_{i\ j}
\right)-\xi^\mathrm{in}\left(^{j\ i}_{i\ j}\right)\right]}
{1-\sum_\rho\left[\lambda_\rho\left(^{j\ j}_{i\ i}\right)-\lambda_\rho\left(^{j\ i}_{i\ j}\right)\right]}\ .
\end{equation}
The denominator comes from the normalization factor which is not exactly 1 due to fermion exchanges between cofermions. The numerator contains direct and exchange processes similar to the ones we found for cobosons [10]. Note that the fermionic nature of the particles, which leads to $F_i^\dag F_j^\dag|v\rangle=-
F_j^\dag F_i^\dag|v\rangle$, induces a minus sign in the $(i\leftrightarrow j)$ permutation, which does not exist in the case of cobosons. The scattering $\xi\left(^{n\ \,j}_{m\ i}\right)$ corresponds to the direct interaction process appearing in Eq.(17) and represented in Fig.6(b), the cofermions $i$ and $j$ keeping their three fermions. By contrast, the scattering 
$\xi^\mathrm{in}\left(^{n\ \,j}_{m\ i}\right)$ corresponds to an exchange interaction process, cofermions $i$ and $j$ exchanging just one fermion --- since a two-fermion exchange is equivalent to a one-fermion exchange with an index permutation. This exchange scattering is defined in the same way as for cobosons [10], namely, 
\begin{equation}
\xi^\mathrm{in}\left(^{n\ \,j}_{m\ i}\right)=\sum_{p,q}\sum_{\rho}\lambda_\rho\left(^{n\ \,q}_{m\ p}\right)
\xi\left(^{q\ j}_{p\ i}\right)\ ;
\end{equation}

\section{Conclusion}

Through a rather intuitive analysis of the fermion exchanges taking place between composite quantum particles, we have identified the possible topologies of the diagrams representing these exchanges. In addition to diagrams similar to the Shiva diagrams introduced to visualize the many-body physics of cobosons made of two fermions, we here show that a new set of ``three-arm'' diagrams, called Kali, is necessary to properly represent fermion exchanges taking place between cofermions made of three fermions.

We also show how these fermion exchanges can be generated from hard algebra, through a set of commutators and anticommutators similar to the ones we introduced in the many-body theory of cobosons made of two fermions. The complexity of these exchanges increasing rapidly with the number of fermions contained in these composite objects, their visualization through Shiva and Kali diagrams, will appear as highly valuable to control the many-body physics of such composite quantum particles.

\vspace{1cm}

We wish to thank Marc-Andr\'{e} Dupertuis for valuable discussions at the beginning of this work.

\clearpage

\begin{figure}
\vspace{-4.5cm}
\centerline{\scalebox{0.5}{\includegraphics{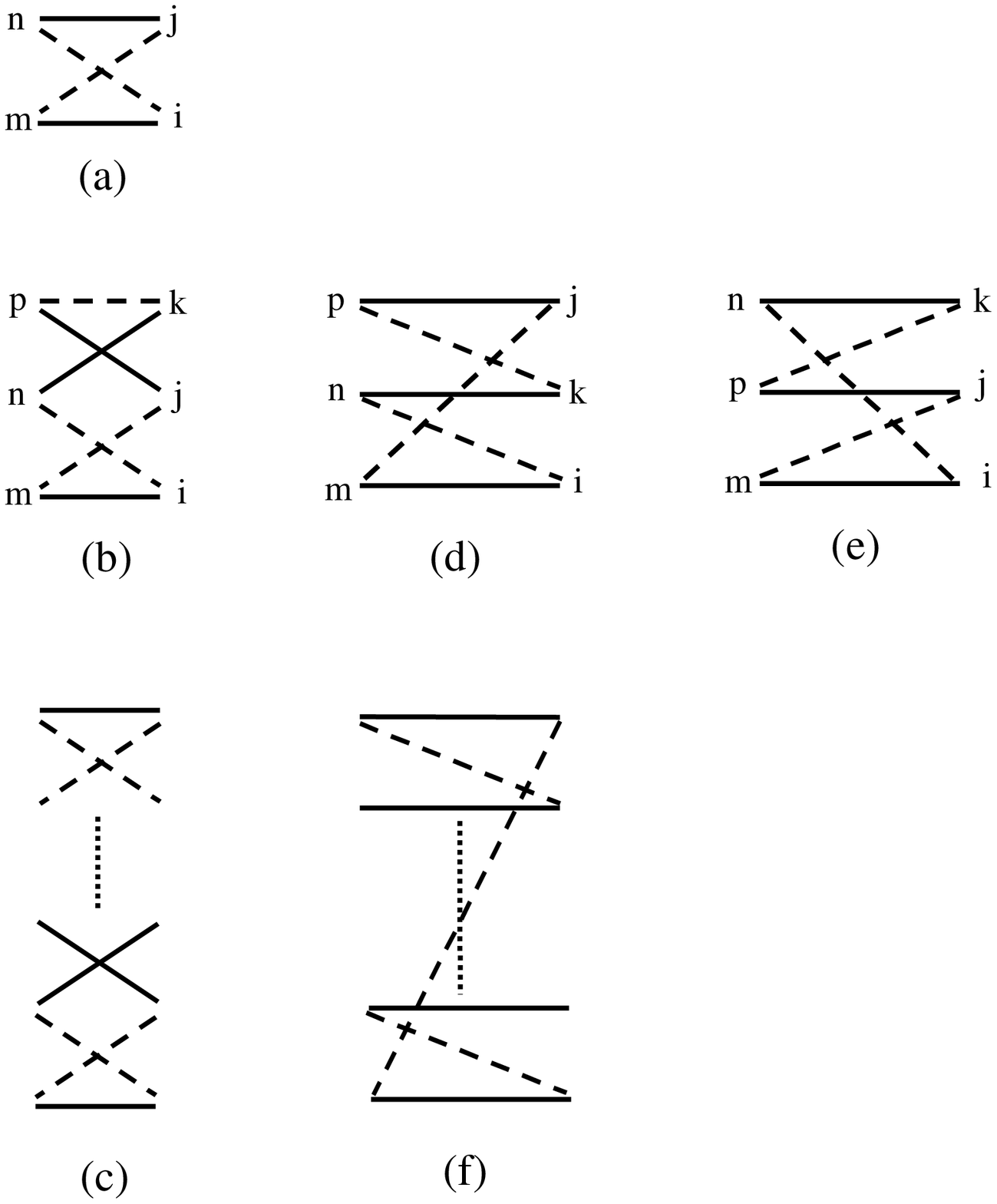}}}
\vspace{-3,5cm}
\caption{Usual ``star-topology'' for Shiva diagrams representing fermion exchanges between two (a), three (b), and $N$ cobosons (c) made of two fermions $\alpha$ (solid lines) and $\beta$ (dashed lines). These exchanges can also be represented through ``line-topology'' as shown in (d,e) or (f).}
\end{figure}

\begin{figure}
\vspace{-1,5cm}
\centerline{\scalebox{0.5}{\includegraphics{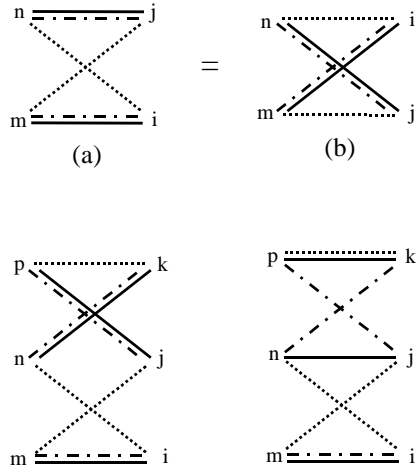}}}
\vspace{-5.5cm}
\caption{Shiva-like diagrams for fermion exchanges between cofermions made of three fermions $(\alpha,\beta,\gamma)$ represented by solid, dashed and dotted lines. In (a,b), the two cofermions explode into (2+1) fermions. In (c), the three cofermions explode into (2+1) fermions, while in (d), only two cofermions explode into (2+1) fermions.}
\end{figure}

\clearpage

\begin{figure}
\vspace{-5cm}
\centerline{\scalebox{0.5}{\includegraphics{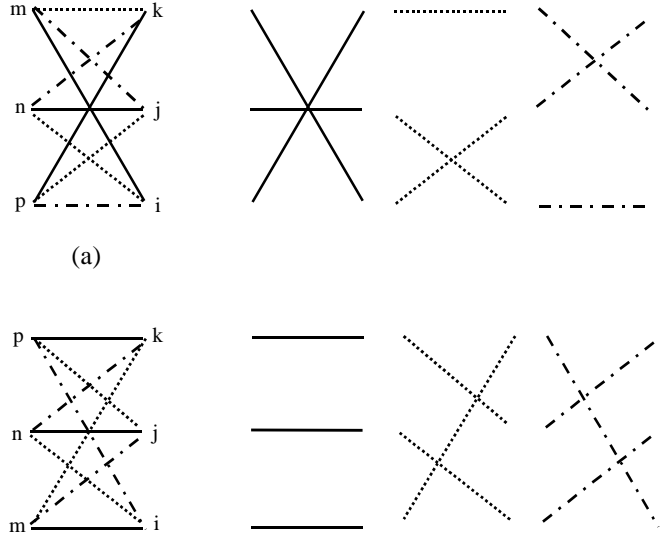}}}
\vspace{-5cm}
\caption{Kali diagrams for fermion exchanges between three cofermions which explode into (1+1+1) fermions. The same exchange process can be equivalently represented in a ``star-topology'' (a) or in a ``line topology'' (b). We have also represented the $(\alpha,\beta,\gamma)$ lines separately, to make the diagram topologies clearer.}
\end{figure}

\begin{figure}
\vspace{-6.5cm}
\centerline{\scalebox{0.5}{\includegraphics{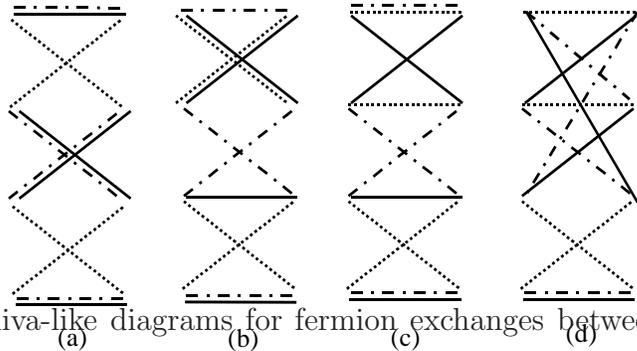}}}
\vspace{-8cm}
\caption{(a,b,c) Shiva-like diagrams for fermion exchanges between four cofermions in which four, three, two cofermions explode into (2+1) fermions. (d) Mixture of Shiva-Kali diagram when only one of the four cofermions explodes into (2+1) fermions.}
\end{figure}

\clearpage

\begin{figure}
\vspace{-7cm}
\centerline{\scalebox{0.5}{\includegraphics{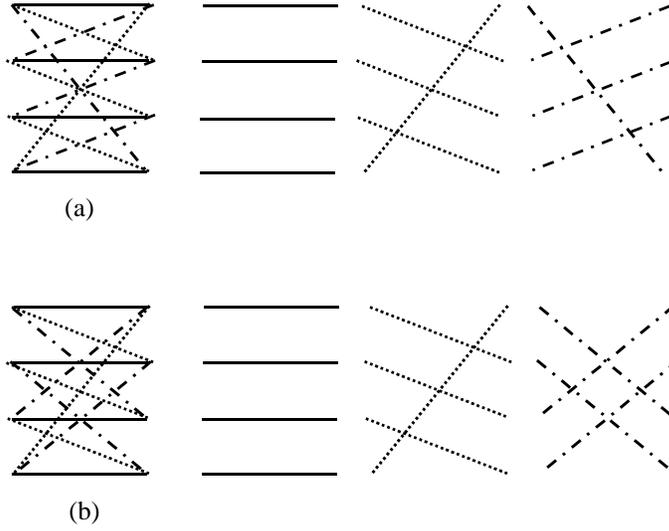}}}
\vspace{-5cm}
\caption{The two possible Kali diagrams in a ``line-topology'' for fermion exchanges between four cofermions exploding into (1+1+1) fermions. The $(\alpha,\beta,\gamma)$ lines, written separately, help to see the topology of these two different diagrams.}
\end{figure}

\begin{figure}
\vspace{-7cm}
\centerline{\scalebox{0.5}{\includegraphics{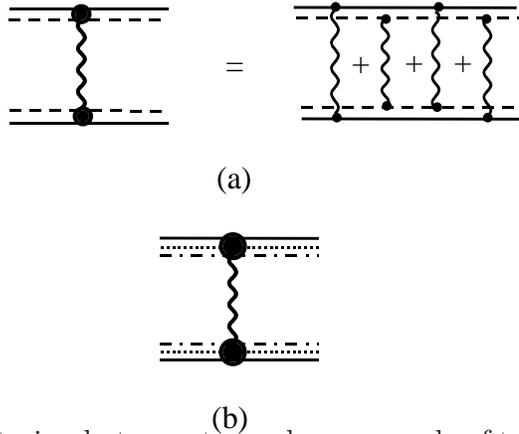}}}
\vspace{-5cm}
\caption{(a) Direct scattering between two cobosons made of two fermions, resulting from interactions between their fermionic components. (b) Same for two cofermions made of three fermions.}
\end{figure}

\end{document}